\documentclass[12pt]{article}
\usepackage{amsmath}
\usepackage{cite}
\usepackage{epsfig}
\usepackage{epstopdf}
\usepackage{amssymb}

\def\[{\left [}
\def\]{\right ]}
\def\({\left (}
\def\){\right )}

\def\r{\rho}

\def\r2{\sqrt{2}}

\newcommand{\bbibitem}[1]{\bibitem{#1}\marginpar{#1}}

\def\Label#1{\label{#1}%
  \smash{\hbox to0pt{\raise1ex\hbox{\tiny[#1]}\hss}}}
\def\noLabels{\let\Label=\label}
\def\nobbibitem{\let\bbibitem=\bibitem}

\newcommand{\bea}{\begin{eqnarray}}
\newcommand{\eea}{\end{eqnarray}}

\newcommand{\beq} {\begin{equation}}
\newcommand{\eeq} {\end{equation}}
\newcommand{\beqa} {\begin{eqnarray}}
\newcommand{\eeqa} {\end{eqnarray}}

\newcommand{\beqn}{\begin{eqnarray}}
\newcommand{\eeqn}{\end{eqnarray}}

\begin{document}

\begin{flushright}
Nordita-2012-28\\
RH-03-2012
\end{flushright}

\vskip 2cm \centerline{\Large {\bf Thermal Correlators in Holographic Models}}
\vskip .5cm \centerline{\Large {\bf with Lifshitz scaling}}
\vskip 2cm
\renewcommand{\thefootnote}{\fnsymbol{footnote}}
\centerline
{{\bf Ville Ker\"{a}nen,$^{1,2}$
\footnote{vkeranen@nordita.org}, Larus Thorlacius,$^{1,2}$
\footnote{larus@nordita.org}
}}
\vskip .5cm
\centerline{\it
${}^{1}$Nordita}
\centerline{\it KTH Royal Institute of Technology and Stockholm University}
\centerline{\it Roslagstullsbacken 23, SE-106 91 Stockholm, Sweden}

\vskip .5cm
\centerline{\it
${}^{2}$ University of Iceland, Science Institute}
\centerline{\it Dunhaga 3, IS-107 Reykjavik, Iceland}

\setcounter{footnote}{0}
\renewcommand{\thefootnote}{\arabic{footnote}}

\begin{abstract}
We study finite temperature effects in two distinct holographic models that
exhibit Lifshitz scaling, looking to identify model independent features in
the dual strong coupling physics. We consider the thermodynamics of black
branes and find different low-temperature behavior of the specific heat.
Deformation away from criticality leads to non-trivial temperature dependence
of correlation functions and we study how the characteristic length scale in the
two point function of scalar operators varies as a function of temperature and
deformation parameters.
\end{abstract}

\section{Introduction}

A number of holographic models have been introduced to
study quantum critical systems in low dimensions,
see \cite{Hartnoll:2009sz,Herzog:2009xv,Horowitz:2010gk,
McGreevy:2009xe,Sachdev:2010ch} for reviews.
In the holographic approach, relatively simple classical or semi-classical
calculations are carried out in a suitably chosen gravitational background
and the results interpreted in terms of strongly coupled physics in the
dual field theory. For condensed matter applications, the holographic
models are usually of the bottom up variety, where key symmetries
and conserved charges of the dual system under study are realized by
coupling gravity to a minimal number of additional fields. Without the
benefit of extended supersymmetry and the pedigree of a consistent
embedding into string theory, the validity of the gauge theory/gravity
correspondence is very much an open question for bottom up models.
On the other hand, they offer a simple phenomenological setting in
which to study a range of physical effects and one hopes to identify
a set of generic features in bottom up models that may also be shared
by more realistic top down constructions. Much of the work in this
area has focused on systems with an underlying conformal symmetry.
This is partly because conformal symmetry is realized in a number of
interesting physical systems, but also because the assumption of
conformal symmetry often simplifies calculations. On the other hand,
conformal symmetry is usually not present in condensed
matter systems and generic quantum critical points exhibit scale
invariance without conformal invariance.

In a quantum phase transition, the ground state of a system changes
abruptly when some parameter of the system is varied through a
critical value (for reviews, see {\it e.g.} \cite{Sachdev,Sondhi}). As in a
conventional phase transition, characteristic
length scales of the system diverge at a quantum critical point and
there is an emergent scaling symmetry of the form
\beq
t\rightarrow \lambda^z\,t , \qquad \vec{x}\rightarrow \lambda \vec{x},
\label{eq:anisotropic}
\eeq
where $z\geq1$ is referred to as the dynamical critical exponent. For
$z>1$ the scaling at the quantum critical point is anisotropic between
the temporal and spatial directions, commonly referred to as Lifshitz
scaling in the literature.

In this paper we study finite temperature effects in two different bottom up holographic
models that exhibit Lifshitz scaling. On the one hand we consider Einstein-Maxwell gravity
coupled to a massive vector (Proca) field. We refer to this as the EMP model (for
Einstein-Maxwell-Proca). It is based on a model introduced in \cite{Kachru:2008yh} involving
a pair of two- and three-form field strengths coupled to Einstein gravity with a negative
cosmological constant. This was reformulated in \cite{Taylor:2008tg} as Einstein gravity
coupled to a Proca field, and in \cite{Brynjolfsson:2009ct} a $U(1)$
Maxwell field was added in order to study dual systems at finite charge density.
The second model was introduced in \cite{Tarrio:2011de} and consists of Einstein gravity
coupled to a dilaton field and a pair of $U(1)$ gauge fields. It is a generalization of an
earlier model considered in \cite{Taylor:2008tg} and will be referred to below as the
EDM model (for Einstein-Dilaton-Maxwell).
A motivation for studying two models at once is to get a handle on how much of the
physics is controlled by the anisotropic scaling invariance (\ref{eq:anisotropic}) and
how much is model dependent.

We compare charged black brane solutions and some of their thermodynamics
between the two models. We then consider two point correlation functions of operators
dual to a probe scalar field in these holographic models. The two point correlators
exhibit screening at finite temperature and we obtain the dependence of the screening
length on temperature and deformation parameters in the two holographic models.
These results can be compared at a qualitative level to corresponding computations
at weak coupling.
In particular, the quantum Lifshitz model provides a well controlled theoretical setting
to study thermal effects near a $z=2$ fixed point \cite{Viswanath,Fradkin,Ardonne}.
The temperature dependence of characteristic length scales was studied via the two
point function of a scaling operator in this model in \cite{Ghaemi}. It was found
that the two point correlation function at spatially separated points vanishes at finite
temperature due to an infrared divergence. This makes the system very sensitive to
deformations away from the critical point. The authors of \cite{Ghaemi} considered
the effect on the two point function of  introducing an irrelevant operator and found
that it restores correlations between spacelike separated points and introduces
a temperature dependent characteristic length scale into the two point function.

In general, heating up a system at a quantum
critical point introduces a length scale that breaks the scaling symmetry,
but at the same time it broadens the region in parameter space that is
controlled by the underlying scale invariant theory.
In an otherwise scaling symmetric theory, the dependence of a characteristic
length scale $\xi$ on temperature is given by
\beq
\xi=c T^{-1/z},\label{eq:trivial}
\eeq
where $c$ is some constant numerical coefficient. As one deforms the system
away from the critical point, the temperature dependence of the characteristic length
scale becomes more generic. Consider a set of deformation parameters $\lambda_i$,
which for convenience are defined so that they have dimensions of inverse length.
At zero temperature $\lambda_i=0$ corresponds to infinite correlation length.
Away from $\lambda_i=0$, the temperature dependence becomes
\beq
\xi=T^{-1/z} \eta(T^{-1/z}\lambda_i),\label{eq:deformed}
\eeq
where $\eta$ is some generic function of its arguments.
The only requirement is that $\eta(0)=c$.

In the holographic models, turning on a finite temperature leads to the exponential
decay of two point correlation functions between spacelike separated points, with
a temperature dependent correlation length $\xi$. If the finite temperature is the
only deformation parameter, then the temperature dependence of the correlation
length will be of the simple form (\ref{eq:trivial}). More interesting behavior is seen
if we consider deformations away from the fixed point. We study two different types
of relevant deformations of the model. The first one is to turn on a chemical potential
for a conserved $U(1)$ charge,
\beq
S_1=\int d\textbf{x}d\tau \mu \rho,
\eeq
which on the gravity side corresponds to considering gravitational solutions with an
electric flux. The relevant black brane solutions in the EMP model were constructed
in \cite{Brynjolfsson:2009ct} and in the EDM model in \cite{Tarrio:2011de}.
As the temperature is lowered, this deformation will take the system from the Lifshitz
fixed point to a "locally critical" fixed point corresponding to the appearance of an
$AdS_2\times R^2$ region in the spacetime.

The second type of deformation we consider is to add a double trace deformation
for the probe scalar field
\beq
S_2=\int d\textbf{x}d\tau\lambda\mathcal{O}^2,
\eeq
whose correlation functions we are interested in. This deformation is less dramatic
than the previous one in that, as the temperature is lowered, the theory flows to
another Lifshitz symmetric fixed point.

The paper is organized as follows. In section \ref{sec:quantumlifs} we briefly review
the calculation of the finite temperature correlation function in the quantum Lifshitz
model in \cite{Ghaemi}. In section \ref{sec:hololifs} we introduce the two holographic
models with Lifshitz scaling and review properties of their charged black brane solutions.
In section \ref{sec:thermo} we compare some thermodynamic properties of black branes
and conclude that the low temperature physics differs between the two models.
In section \ref{sec:densitycorr} we obtain two point correlation functions in a
charged black brane background. We first consider operators of large scaling
dimensions, for which the two point correlator can be obtained in a geodesic
approximation, and then repeat the analysis for operators of general scaling
dimensions. In section \ref{sec:doubletracecorr} we calculate the two point correlation
function with a double trace deformation at finite temperature (and at a vanishing
charge density). Finally in section \ref{sec:conclusions} we summarize our results.

\section{Thermal correlators in the quantum Lifshitz model}\label{sec:quantumlifs}

A free field theory realization of a class of $z=2$ quantum critical points is provided
by the quantum Lifshitz model,
\beq
S_0=\frac{1}{2}\int d^2xd\tau\Big((\partial_{\tau}\chi)^2+K(\nabla^2\chi)^2\Big).
\eeq
Here we will briefly review the finite temperature correlation functions in this model,
obtained in \cite{Ghaemi}, in order to compare them to the results of holographic
calculations later in the paper. We refer to \cite{Viswanath,Fradkin,Ardonne} for
more detailed discussions of the quantum Lifshitz model.

The operators of interest are the so called monopole operators\footnote{The name
monopole operator comes from a dual gauge theory representation of the same
model, where one has $E_i=\epsilon_{ij}\partial_j\chi$. The operator in
(\ref{eq:monopole}) creates a monopole in the dual gauge field \cite{Ghaemi}.}
\beq
\mathcal{O}(x)=e^{2\pi i\chi(x)}.\label{eq:monopole}
\eeq
The vacuum two point correlation function has the scaling form
\beq
G(\textbf{x},\textbf{x}')=
\langle\mathcal{O}(\textbf{x},0)\mathcal{O}(\textbf{x}',0)\rangle
\propto \frac{1}{|\textbf{x}-\textbf{x}'|^{\pi/\sqrt{K}}},\label{eq:vacuumscaling}
\eeq
which follows from logarithmic correlations of $\chi$. At finite temperature the
contribution from the zeroth Matsubara mode to the $\chi$ correlator is proportional to
\beq
T\int d^2k\frac{1-e^{i\textbf{k}\cdot(\textbf{x}-\textbf{x}')}}{K\textbf{k}^4},
\eeq
where the momentum integral is seen to diverge logarithmically with an infrared
cutoff $L$. The two point correlation function of spacelike separated monopole
operators thus vanishes at finite temperature in the infinite volume limit,
\beq
G(\textbf{x},\textbf{x}')\propto \frac{1}{|\textbf{x}-\textbf{x}'|^{\pi/\sqrt{K}}}
e^{-\frac{\pi T}{2K}|\textbf{x}-\textbf{x}'|^2\log L}\rightarrow 0.
\eeq
Spatial correlations are restored away from the scaling limit of the free theory.
The authors of \cite{Ghaemi} consider a marginally irrelevant deformation of the
free model, by adding an interaction term
\beq
S_{int}=\frac{1}{2}\int d^2xd\tau u(\nabla\chi)^4,
\eeq
which corresponds to slightly moving away from the critical point. This deformation
leads to a logarithmic violation of scaling symmetry, which changes the finite
temperature behavior. The interaction induces a term
\beq
v^2(\nabla\chi)^2,
\eeq
into the effective action for $\chi$ already at one loop. This term changes the low
momentum behavior of the theory dramatically and cures the infrared divergences.
By using a large $N$ gap equation the authors of \cite{Ghaemi} find
\beq
v^2\approx 4T\sqrt{K}\frac{\log\Big(-\log T\Big)}{-\log T},
\eeq
which is understood to hold for small $T$. This means that there is a non-trivial
(temperature dependent) length scale induced in the theory
\beq
\xi^2_T=\frac{K}{v^2}.\label{eq:QLscale}
\eeq
Using the new effective action leads to a two point correlator of monopole operators
of the following form:
\beq
G(\textbf{x},\textbf{x}')\propto
\exp \left[-\frac{\pi T}{2 v^2}\frac{|\textbf{x}-\textbf{x}'|^2}{\xi_T^2}
\left(\log{\left[\frac{2\xi_T}{|\textbf{x}-\textbf{x}'|}\right]}+c\right)\right],
\eeq
where $c$ is a constant, for distances smaller than $\xi_T$,  and
\beq
G(\textbf{x},\textbf{x}')\propto
\exp{\left[-\frac{2\pi T}{v^2}\log\left(\frac{|\textbf{x}-\textbf{x}'|}{\xi_T}\right)\right]},
\eeq
for distances larger than $\xi_T$.
Furthermore, if one allows for vortices (spinons) in the $\chi$ field, the correlation
function changes qualitatively at very large distances. Assuming a vortex energy
gap $E_c$, one finds that correlations are exponentially decaying on a length scale
$\xi_{vortex}\propto e^{E_c/2T}$ due to a vortex plasma \cite{Ghaemi}.

To summarize, the authors of \cite{Ghaemi} find a non-trivial temperature dependence
in correlation functions of spatially separated monopole operators, which arises on the
one hand from the Lifshitz scaling region (at short distances) and on the other hand from
deformations away from criticality.
It seems likely that many of the details depend on the specific theory and the specific
deformations that were considered in \cite{Ghaemi}.
In sections~\ref{sec:densitycorr} and~\ref{sec:doubletracecorr} below, we calculate
two point correlation functions of scaling operators in holographic models exhibiting
Lifshitz scaling. We also find a temperature dependent thermal length scale which can
be compared to that of the quantum Lifshitz model. Since the holographic models are
dual to strongly coupled theories and the quantum Lifshitz model is a free field theory,
one does not expect to reproduce the results of \cite{Ghaemi} in detail but rather look
whether the holographic models exhibit similar trends.

\section{Holographic models with Lifshitz scaling}\label{sec:hololifs}
We will consider two different holographic models, both of which realize anisotropic
scaling of the form (\ref{eq:anisotropic}) through spacetime geometries that are
asymptotic to the so-called Lifshitz geometry \cite{Kachru:2008yh,Koroteev:2007yp},
\beq
ds^2=\ell^2\left(-r^{2z}dt^2+\frac{dr^2}{r^2}+r^2d\textbf{x}^2\right).
\label{eq:lifshitzgeometry}
\eeq
Here $\ell$ is a characteristic length scale of the geometry and the
$r$, $t$, and $\vec{x}$ coordinates have no length dimensions.
The Lifshitz metric is invariant under the transformation
\beq
t\rightarrow \lambda^z\,t , \qquad \vec{x}\rightarrow \lambda \vec{x} , \qquad
r \rightarrow \frac{r}{\lambda} ,
\label{eq:lifshitzscaling}
\eeq
which incorporates the scaling in (\ref{eq:anisotropic}) on the $t$ and $\vec{x}$
coordinates, while the radial coordinate $r$ of the bulk geometry scales inversely
with $\lambda$. In the rest of the paper the characteristic length will be set to $\ell=1$.
It can be re-introduced via dimensional analysis if needed.
For concreteness, we will take the bulk spacetime to be 3+1 dimensional, but
our results generalize in a straightforward fashion to other dimensions.

The two models both involve Einstein-Maxwell gravity with a negative cosmological
constant but the asymptotic Lifshitz behavior is achieved by coupling to different matter
sectors. Both models have charged black brane solutions that are asymptotic to the
Lifshitz geometry and are interpreted as holographic duals of field theory configurations
at finite temperature and finite charge density. The black brane geometries are not the
same in the two models. In particular, their near-extremal limits differ and this has
consequences for low-temperature physics in the corresponding dual field theories.
One of the aims of this paper is to compare predictions for physical observables at
finite temperature and explore to what extent they are model dependent.

\subsection{Einstein-Maxwell-Proca theory}
The first model we consider is a modified version of the holographic model first
considered in \cite{Kachru:2008yh}. In the original version, anisotropic scaling of the
form (\ref{eq:lifshitzscaling}) was obtained by coupling Einstein gravity to a pair of
two- and three-form field strengths. As was shown in \cite{Taylor:2008tg}, this can
be rewritten as a single massive vector field coupled to gravity and we find this
formulation more convenient to work with. Finally, following \cite{Brynjolfsson:2009ct}, we
include a Maxwell gauge field which is coupled to the gravitational field but not directly
to the auxiliary massive vector field,
\beq
S_\textrm{EMP} = \int\mathrm{d}^4x\sqrt{-g}\;\left(
R-2\Lambda-\frac{1}{4}F_{\mu\nu}F^{\mu\nu}
-\frac{1}{4} \mathcal{F}_{\mu\nu}\mathcal{F}^{\mu\nu}
-\frac{c^2}{2}\mathcal{A}_\mu\mathcal{A}^\mu\right) .
\label{eq:action1}
\eeq
The role of the massive vector fieldÊ background is to modify the asymptotic behavior
of the metric. One could Êin principle includeÊ direct couplings between the massive
vector and the Maxwell gauge field, in which case having a massive vector background
field would also influence the gauge field dynamics, but for simplicity we have chosen
not to do that. In what follows we will refer to this theory as the EMP model.

The Lifshitz metric (\ref{eq:lifshitzgeometry}) is a solution of the equations of motion
when the constants in the model are related to the dynamical critical exponent $z$
through
\beq
\Lambda=- \frac{1}{2}(z^{2}+z+4) , \qquad c=\sqrt{2z},
\label{eq:parameters1}
\eeq
and the massive vector field has a non-vanishing background value,
\beq
\mathcal{A}_t=\sqrt{\frac{2(z{-}1)}{z}} r^z \,, \qquad
\mathcal{A}_{x_i}=\mathcal{A}_r=0 \,.
\label{eq:procabackground}
\eeq
The field equations of the EMP model have been studied extensively by many authors
(see \cite{Brynjolfsson:2010mk} for a brief review) and we will not repeat
the analysis here. For generic values of the dynamical critical exponent,
there are no known analytic solutions describing black holes or black
branes\footnote{An exact solution exists for $z=4$ (or more
generally $z=2d$ where $d$ is the number of transverse spatial dimensions)
and a particular value of the electric charge on the black hole, or charge
density on the black brane \cite{Brynjolfsson:2009ct,Pang:2009pd}.}
but numerical solutions can easily be found using similar techniques
as introduced in \cite{Danielsson:2009gi}.  Below, we will present results
obtained from numerical solutions of the EMP model and compare them
with results obtained using analytic black brane solutions from the other model.

\subsection{Einstein-Dilaton-Maxwell theory}
The second model is an Einstein-Dilaton-Maxwell (EDM) theory
\beq
S_\textrm{EDM} = \int\mathrm{d}^4x\sqrt{-g}\;\left[
R-2\Lambda-\frac{1}{2}\partial_\mu\phi\partial^\mu\phi
-\frac{1}{4}\sum_{i=1}^2 e^{\lambda_i\phi}F^{(i)}_{\mu\nu}F^{(i)\mu\nu}\right].
\label{eq:action2}
\eeq
Models of this type were introduced in the context of non-relativistic
holography in \cite{Taylor:2008tg} and black brane solutions of this particular
action were found in \cite{Tarrio:2011de}. Dynamical solutions describing
infalling energy were constructed in \cite{Keranen:2011xs} and used to study
thermalization following a non-relativistic holographic quench but here
we will only consider static solutions.

The equations of motion are
\beqa
R_{\mu\nu}-\frac{1}{2}R g_{\mu\nu}+\Lambda g_{\mu\nu}&=&T^\phi_{\mu\nu}
+T^{(1)}_{\mu\nu}+T^{(2)}_{\mu\nu} , \label{edmeinstein}\\
\nabla^2\phi- \sum_{i=1}^2 \frac{\lambda_i}{4} e^{\lambda_i\phi}
F^{(i)}_{\mu\nu}F^{(i)\mu\nu} &=&0 , \label{edmdilatoneq}\\
\nabla_\mu\left(e^{\lambda_i\phi} F^{(i)\mu\nu}\right) &=& 0 , \qquad
i=1,2 \,,  \label{edmmaxwell}
\eeqa
with
\beqa
T^\phi_{\mu\nu}&=&\frac{1}{2}\partial_\mu\phi\partial_\nu\phi
-\frac{1}{4} g_{\mu\nu}(\partial\phi)^2 , \\
T^{(i)}_{\mu\nu}&=&\frac{e^{\lambda_i\phi}}{2}\left[
F^{(i)}_{\mu\sigma}F^{(i)\>\sigma}_{\phantom{()}\nu} -\frac{1}{4}g_{\mu\nu}
F^{(i)}_{\sigma\rho}F^{(i)\sigma\rho}\right] .
\eeqa
Remarkably, these field equations can be solved analytically to obtain charged
black branes in asymptotically Lifshitz spacetime for any value of the dynamical
critical exponent $z$. Below we reproduce the static black brane solutions
of \cite{Tarrio:2011de} and review some of their thermodynamic properties. We
then use them to calculate finite temperature correlation functions of scalar
operators in the dual field theory. At each stage, we compare our results to
corresponding calculations in the EMP model.

\subsubsection{Asymptotically Lifshitz solutions}
We use the following metric ansatz for static solutions,
\beq
ds^2= -r^{2z}f(r) dt^2+\frac{dr^2}{r^2 g(r)}+r^2(dx^2+dy^2) ,\label{eq:metricansatz}
\eeq
for which the field equations (\ref{edmmaxwell}) for the gauge fields are solved by
\beq
F^{(i)}_{rt}=\rho_i\, r_0^{z-1}\left(\frac{r}{r_0}\right)^{z-3}
\sqrt{\frac{f(r)}{g(r)}} \,e^{-\lambda_i\phi(r)} ,
\label{solvemaxwell}
\eeq
where $r_0$ is an arbitrary reference value of the radial variable and the
$\rho_i$ are dimensionless constants. Under the coordinate rescaling
(\ref{eq:lifshitzscaling}), $r_0$ transforms like $r$ while $\rho_i$ remain invariant.
Later on, when we consider black hole solutions it will be
convenient to take $r_0$ as the radial location of the event horizon.

The Einstein equations (\ref{edmeinstein}) reduce to a pair of first order
differential equations,
\beqa
\frac{rf'}{f}-\frac{rg'}{g} &=&-2(z-1)+\frac{1}{2}(r\phi')^2 , \label{einstein1}\\
\frac{rf'}{f}+\frac{rg'}{g} &=&-2(z+2)-\frac{1}{g}\left[2\Lambda
+\frac{1}{2}\left(\frac{r_0}{r}\right)^4\left(
\rho_1^2e^{-\lambda_1\phi} +\rho_2^2e^{-\lambda_2\phi}\right) \right] ,\label{einstein2}
\eeqa
and the dilaton equation (\ref{edmdilatoneq}) becomes
\beq
r^2 \phi''+\left[\frac{r}{2}\frac{f'}{f}+\frac{r}{2}\frac{g'}{g}+z-3\right]r\phi'
+\frac{1}{2g}\left(\frac{r_0}{r}\right)^4\left(
\lambda_1\rho_1^2e^{-\lambda_1\phi}
+\lambda_2\rho_2^2e^{-\lambda_2\phi}\right)=0 . \label{dilatoneq}
\eeq
The Lifshitz geometry (\ref{eq:lifshitzgeometry}) has $f(r)=g(r)=1$, for which
equation (\ref{einstein1}) is solved by $e^\phi=\mu r^{2\sqrt{z-1}}$. The integration
constant $\mu$ can be rescaled by using a symmetry of the field equations under
a constant shift of $\phi$ accompanied by a compensating rescaling of the $\rho_i$,
and we find it convenient to normalize it as follows,
\beq
e^\phi=\left(\frac{r}{r_0}\right)^{2\sqrt{z-1}} ,
\label{lifshitzdilaton}
\eeq
or equivalently set $\mu\, r_0^{2\sqrt{z-1}}=1$.
To determine the parameters of the Lifshitz background we insert this into the remaining
field equations and set $\rho_2=0$. From the dilaton equation (\ref{dilatoneq}) we obtain
\beq
2(z+2)\sqrt{z-1}=-\frac{\lambda_1\rho_1^2}{2}
\left(\frac{r_0}{r}\right)^{4+2\sqrt{z-1}\lambda_1} ,
\eeq
which requires
 \beq
 \lambda_1=-\frac{2}{\sqrt{z-1}} , \qquad
\rho_1=\sqrt{2(z-1)(z+2)} ,
\eeq
and equation (\ref{einstein2}) then relates the cosmological constant
to the dynamical critical exponent,
 \beq
 \Lambda=-\frac{1}{2}(z+1)(z+2) .
\eeq
We note that $\lambda_1$ is negative. The sign of $\lambda_i$ in (\ref{eq:action2})
determines how the coupling of the corresponding gauge field changes with radial
location and with the dilaton profile in (\ref{lifshitzdilaton}) we find that $F^{(1)}_{\mu\nu}$
is strongly coupled in the asymptotic $r\rightarrow\infty$ region. This makes the model
less satisfactory, but it is not a serious problem as long as $F^{(1)}_{\mu\nu}$ does not
couple directly to anything outside the gravitational sector. Its only role should
be to modify the asymptotic behavior of the metric from AdS to Lifshitz through the
gravitational back reaction to its background value. For applications, such as holographic
superconductors or fermion spectral densities, any additional matter fields in the model
should only be charged under $F^{(2)}_{\mu\nu}$ and not under $F^{(1)}_{\mu\nu}$.

\subsubsection{Charged black branes in EDM model}
The charged black brane solutions in \cite{Tarrio:2011de} have $f=g$ and the
same dilaton profile (\ref{lifshitzdilaton}) as in the purely Lifshitz background.
The Einstein equation (\ref{einstein2}) then reduces to
\beq
 r g'+(z+2)(g-1)=-\frac{\rho_2^2}{4}
 \left(\frac{r_0}{r}\right)^{4+2\sqrt{z-1}\lambda_2} ,
\eeq
while the dilaton equation (\ref{dilatoneq}) is satisfied if
\beq
 r g'+(z+2)(g-1)=-\frac{\lambda_2\rho_2^2}{4\sqrt{z-1}}
 \left(\frac{r_0}{r}\right)^{4+2\sqrt{z-1}\lambda_2} .
\eeq
It follows that $\lambda_2=\sqrt{z-1}$ and the solution for a charged black brane
with an event horizon at $r=r_0$ is given by
\beqa
f(r)\ =\ g(r)&=&1-\left(1+\frac{\rho_2^2}{4z} \right) \left(\frac{r_0}{r}\right)^{z+2}
+\frac{\rho_2^2}{4z} \left(\frac{r_0}{r}\right)^{2z+2} , \label{eq:branemetric} \\
F^{(1)}_{rt}&=&\sqrt{2(z-1)(z+2)}\, r_0^{z-1} \left(\frac{r}{r_0}\right)^{z+1} , \\
F^{(2)}_{rt}&=&\rho_2\, r_0^{z-1} \left(\frac{r_0}{r}\right)^{z+1} , \label{eq:braneF2} \\
e^\phi&=&\left(\frac{r}{r_0}\right)^{2\sqrt{z-1}} \label{eq:branephi} ,
\eeqa
This solution is valid for any value of $z\geq 1$ and can be generalized in a
straightforward fashion to more general spacetime dimensions \cite{Tarrio:2011de}.
In the $z\rightarrow 1$ limit, the metric and the Maxwell gauge field $F^{(2)}_{rt}$ reduce
to those of a standard AdS-Reissner-Nordstr\"om black brane, while the auxilliary gauge
field $F^{(1)}_{rt}$ vanishes and the dilaton becomes independent of $r$.

The scale transformation (\ref{eq:lifshitzscaling}) is a symmetry of the black brane metric
(\ref{eq:branemetric}), provided $r_0$ is rescaled in the same way as $r$.
This is a special feature of black branes with a planar horizon and
it implies that the radial location of the horizon $r_0$ does not have any physical meaning
by itself. The same is true for planar black branes in the EMP model. It is therefore important
to always use scale invariant combinations of physical quantities when presenting results
obtained in these holographic models.

\subsubsection{Asymptotic behavior}
A distinctive feature of numerical black brane solutions in the EMP model is the
presence of a mode that is not well behaved in the $r\rightarrow\infty$ asymptotic
region \cite{Danielsson:2009gi,Bertoldi:2009vn,Ross:2009ar}. Depending on the
value of the dynamical critical exponent, it can either be a growing mode that takes
the system away from the Lifshitz fixed point or a decaying mode that only converges
very slowly to the asymptotic fixed point. The solutions are usually
obtained by numerically integrating the field equations of the model, starting from
initial conditions near the horizon and proceeding out towards the asymptotic region.
The initial data at the horizon is then fine tuned to remove the mode in question.
This works reasonably well over a range of black brane temperature and charge
density but the required fine tuning becomes progressively more difficult as the
temperature is lowered and the black brane charge approaches its extremal value.

\begin{figure}[h]
\begin{center}
\includegraphics[scale=1.15]{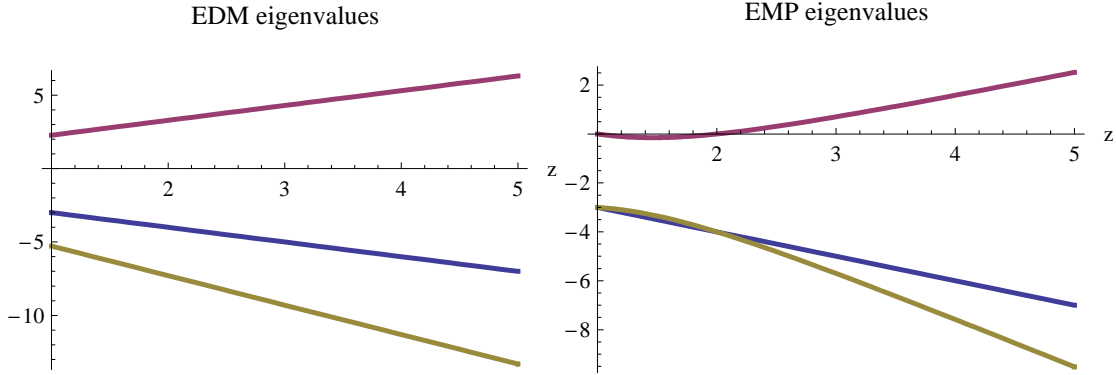}
\caption{\label{fig:eigenvalues} A comparison of mode eigenvalues of the linearized
field equations around a Lifshitz background in the EDM model (left) and the EMP
model (right).}
\end{center}
\end{figure}

In the large $r$ limit, the analytic black brane solutions of the EDM model have
modes that go as $r^{-z-2}$ and $r^{-2z-2}$, and the accompanying coefficients
are the brane energy density and the charge density respectively. It turns out,
however, that the Lifshitz fixed point is always unstable in this model and that
a generic perturbation around it will include a growing mode that is not present
in the black brane solutions.
To see this, we linearize the field equations (\ref{einstein1}) - (\ref{dilatoneq})
around the Lifshitz fixed point. The function $f(r)$ can be eliminated between the
two Einstein equations, leaving a system involving only $g(r)$ and $\phi(r)$.
We allow for a dilaton that varies from its background value,
\beq
e^{\phi(r)} =\left(\frac{r}{r_0} \right)^{2\sqrt{z-1}} \, e^{\sqrt{z-1}\,\varphi(r)} ,
\eeq
and write $g(r)\approx 1+\gamma(r)$. Working to first order in
$\gamma(r)$, $\varphi(r)$, and $\chi(r)\equiv r \varphi'(r)$, we obtain
\beq
r \frac{d}{dr}\left[
\begin{array}{c}
\gamma \\
\varphi \\
\chi
\end{array}
\right]=\left[
\begin{array}{ccc}
-z{-}2 & -(z{-}1)(z{+}2) & -z{+}1 \\
0 & 0 & 1 \\
0 & 2(z{+}1)(z{+}2) & -z{-}2
\end{array}
\right] \left[
\begin{array}{c}
\gamma \\
\varphi \\
\chi
\end{array}\right]-\frac{\rho_2^2}{4}\left(\frac{r_0}{r}\right)^{2z+2}
\left[
\begin{array}{c}
1 \\
0 \\
0
\end{array}\right],
\label{lineareqs}
\eeq
At large $r$, a solution to the full non-linear system will approach a linear
combination of the eigenmodes of the linearized system plus a universal
mode coming from the source term. The eigenvalues,
\beq
-z-2 , \quad \frac{1}{2}\left(-z-2\pm\sqrt{(z+2)(9z+10)}\right) ,
\eeq
are plotted on the left in Figure~\ref{fig:eigenvalues}, showing that the system
has a growing mode for any value of $z$. The result of the corresponding mode analysis
of the EMP model \cite{Bertoldi:2009vn} is shown on the right in the figure, for comparison.
In both models, turning on a growing mode would change the asymptotic behavior
of the metric Êand Êtake the dual field theory away from the Lifshitz UV fixed point.
In this paperÊ we are interested Êin models with Lifshitz scaling and we therefore only
consider configurations Êwhere the growing mode is not turned on.

Both models have an eigenmode that falls off as $r^{-z-2}$ and from a general analysis of
holographic renormalization in asymptotically Lifshitz
spacetime \cite{Zingg:2011cw,Ross:2011gu,Baggio:2011cp,Mann:2011hg,Tarrio:2012xx} one
concludes that the coefficient in front of such a mode is proportional to the energy density
of the configuration. In both models the brane electric charge density enters the linearized
system as a universal mode due to a source term but the falloff with $r$ is different.
In the EMP model the charge mode falls off as $r^{-4}$, independent of $z$, while in the
EDM model the falloff is $r^{-2z-2}$ and at leading order this mode only enters
in the metric function and not in the dilaton field.

\subsubsection{Extremal Lifshitz black branes}
At the extremal value of the charge density,
$\rho_2=\pm \sqrt{4(z+2)}$, the metric function (\ref{eq:branemetric})
in the EDM model has a double zero at the horizon,
\beq
f(r)\approx (z+1)(z+2)\left(\frac{r}{r_0}-1\right)^2 \,
\eeq
and the near horizon metric can be written in $\textrm{AdS}_2\times \textrm{R}_2$ form,
\beq
ds^2\approx -\frac{1}{(z+1)(z+2)}\left[-v^2 d\hat t^2+\frac{dv^2}{v^2}\right]
+d\hat x^2+d\hat y^2 ,\label{eq:ads2metric}
\eeq
where
\beq
v= \frac{r}{r_0}-1, \quad
\hat t =(z+1)(z+2)r_0^z\,t, \quad
\hat x=r_0\, x , \quad \textrm{and}\quad \hat y=r_0\, y .
\eeq
This means that the geometry of extremal Lifshitz black branes in
the EDM model is qualitatively similar to that of an extremal AdS-Reissner-Nordstr\"om
black brane at $z=1$. It follows that various holographic computations that have been
performed in AdS-RN backgrounds will carry over to the EDM model in a straightforward
fashion. In particular, the single fermion spectral function for probe fermions can be
evaluated using similar techniques as have been used in asymptotically AdS
spacetime \cite{Cubrovic:2009ye,Liu:2009dm,Faulkner:2009wj} and we expect
qualitatively similar results, including signals of non Fermi liquid behavior, to be obtained
in the EDM model\footnote{For calculations of fermion correlation functions in Lifshitz
backrounds see \cite{Gursoy:2011gz,Fang:2012pw,Alishahiha:2012nm}.}.

Less is known about extremal black branes in the EMP model.
The commonly used method for finding numerical black brane solutions breaks down
before the extremal limit is reached and the known exact solution at $z=4$ is non-extremal.
An analysis of the field equations of the EMP model
at extremal charge density suggests that the near horizon geometry of an extremal
black brane is in fact $\textrm{AdS}_2\times \textrm{R}_2$ in this model as well.
The near horizon metric can be brought into the same form as (\ref{eq:ads2metric}),
but in this case the required coordinate transformation turns out to be non-analytic at
$r=r_0$ and the Proca field also goes to zero in a non-analytic fashion at $r=r_0$.

\section{Black brane thermodynamics}\label{sec:thermo}
\begin{figure}[h]
\begin{center}
\includegraphics[scale=1.15]{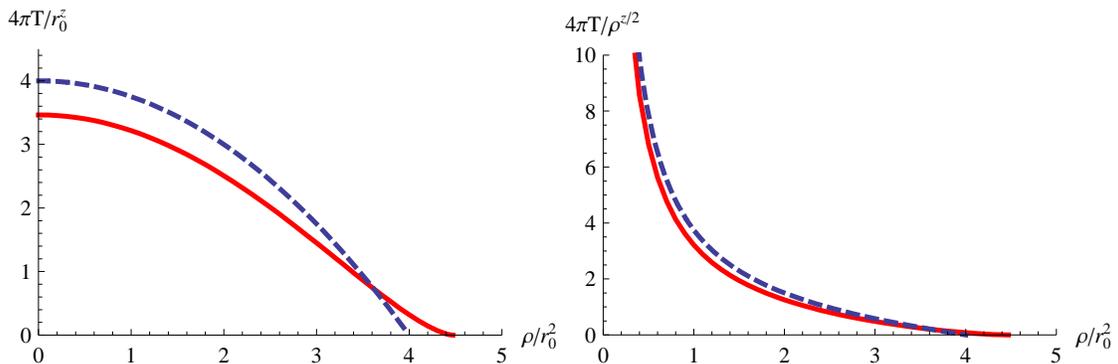}
\caption{\label{fig:temperature} Hawking temperature as a function of the dimensionless charge density for $z=2$ charged black branes. On the left, the dashed (blue) curve shows the
temperature in the EDM model (equation (\ref{eq:edmtemp})) while the solid (red) curve is from
numerical solutions of the EMP model. On the right, the scale invariant combination
$T/\rho$ is shown for the same $z=2$ data.}
\end{center}
\end{figure}

The Hawking temperature of an asymptotically Lifshitz black brane is determined
in the usual way by considering the near horizon behavior of the metric,
\beq
 T=\frac{r_0^{z+1}}{4\pi}\sqrt{ f'(r_0) g'(r_0)} .\label{eq:temperature}
 \eeq
For the analytic black brane solutions of the EDM model one finds
 \beq
 T=\frac{r_0^z}{4\pi}\left[z+2-\frac{\rho_2^2}{4}\right] ,
 \label{eq:edmtemp}
 \eeq
while, as usual, only numerical results are available for the EMP model.
Figure~\ref{fig:temperature} compares the Hawking temperatures of $z=2$ black branes
in the two models. Other values of $z$ give rise to qualitatively similar graphs.

The EDM action (\ref{eq:action2}) does not depend explicitly on
the gauge potential $A^{(2)}_t$ but only its radial derivative.
The associated radially conserved quantity is the charge density carried by
the black brane,
\beq
\rho\equiv\frac{\delta S_\textrm{EDM}}{\delta \partial_r A^{(2)}_t}
=r^{3-z}\sqrt{\frac{g}{f}}e^{\lambda_2\phi}F^{(2)}_{rt} ,
\eeq
and inserting $F^{(2)}_{rt}$ and $e^\phi$ from (\ref{eq:braneF2}) and (\ref{eq:branephi})
gives
\beq
\rho=\rho_2 r_0^2 .
\label{eq:rho}
\eeq
In the following, we have chosen to keep fixed the physical charge density $\rho$
when calculating thermodynamic quantities and express our results in terms of scale
invariant combinations such as $T/\rho^{z/2}$. An alternative choice is to instead keep
fixed the chemical potential of the Maxwell field.

\begin{figure}[h]
\begin{center}
\includegraphics[scale=1.15]{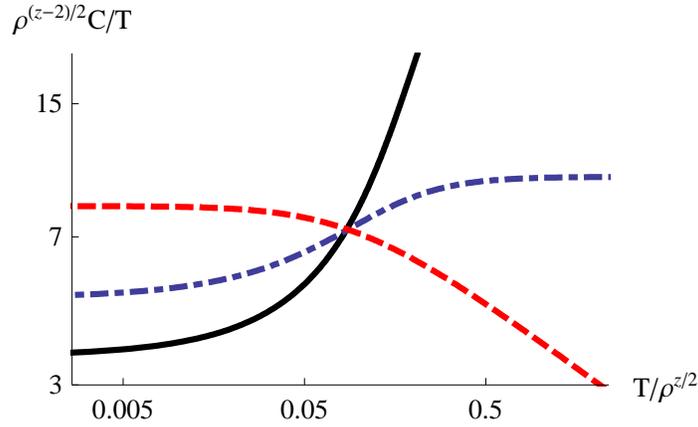}
\caption{\label{fig:specificheat} The ratio of specific heat to temperature in the EDM model for
dynamical scaling exponents $z=1$ (black solid), $z=2$ (blue dot-dashed) and $z=3$ (red dashed).}
\end{center}
\end{figure}

The specific heat at fixed transverse volume and charge density is easily calculated in
the EDM model. It can then be compared to numerical results from the EMP model,
obtained in \cite{Brynjolfsson:2010rx}, and one finds that the two models predict
different low-temperature behavior for this observable. The specific heat at fixed
volume and charge density in the dual boundary theory is given by
\beq
C_{V,\rho}= T\,\frac{dS}{dT},
\eeq
where $S=\pi r_0^2$ is the Bekenstein-Hawking entropy density of the black
brane and the temperature $T$ in the boundary theory is identified with the Hawking
temperature (\ref{eq:temperature}). Keeping $\rho$ in (\ref{eq:rho}) fixed implies
\beq
\frac{d}{dr_0} = \frac{\partial}{\partial r_0}
-\frac{2\rho_2}{r_0}\frac{\partial}{\partial \rho_2},
\eeq
and one then finds
\beqa
\frac{C_{V,\rho}}{T}&=& \frac{dS/dr_0}{dT/dr_0} \nonumber \\
&=& \frac{32\pi^2}{4z(z+2)-(z-4)\rho_2^2}
\left(\frac{\rho}{\rho_2}\right)^{(2-z)/2} . \label{eq:covert}
\eeqa
Figure~\ref{fig:specificheat} plots the scale invariant combination
$\rho^{(z-2)/2}C_{V,\rho}/T$
against $T/\rho^{z/2}$ for three different values of $z$. In all cases the $C/T$ ratio
goes to a constant at low temperature. The low temperature limit can also
be seen directly by going to the extremal limit $\rho_2^2\rightarrow 4(z+2)$ in
the above expression for $C/T$,
\beq
\rho^{(z-2)/2}\,\frac{C_{V,\rho}}{T}
\rightarrow 2^{z/2}(z+2)^{(z-6)/4}\pi^2 \quad \textrm{as}\quad T\rightarrow 0 .
\eeq
The low temperature behavior of the specific heat is very different in the EMP
model. Numerical calculations reported in \cite{Brynjolfsson:2010rx} show
a $C/T$ ratio at fixed charge density that grows as the temperature is lowered
and diverges in the $T\rightarrow 0$ limit for all $z>1$.

At high temperatures the specific heat in (\ref{eq:covert}) satisfies a simple
scaling law consistent with the underlying Lifshitz symmetry,
\beq
C_{V,\rho}\sim T^{2/z} .
\eeq
This scaling is also seen in the EMP model \cite{Brynjolfsson:2010rx}.\footnote{The
specific heat only depends on the geometry (through the entropy) and not directly
on the dilaton field. More general observables, such as correlation functions of fields
coupled to the dilaton, can fail to be Lifshitz symmetric because the dilaton itself
transforms under a Lifshitz rescaling.}
It follows on general grounds from the statistical mechanics of a system
with $\omega\sim k^z$ dispersion in two spatial dimensions \cite{Bertoldi:2009dt}.

\section{Two point functions at finite charge density}\label{sec:densitycorr}

We now turn our attention to finite temperature correlation functions.
For earlier studies of scalar correlators in Lifshitz black hole backgrounds
see \cite{Balasubramanian:2009rx,Giacomini:2012hg}.
In this section we study the temperature dependence of two point correlation functions
of scalar operators at fixed charge density. We first use the geodesic approximation,
valid for large scaling dimensions, to simplify the calculation and get a sense of the
finite temperature behavior.\footnote{The finite temperature two point function at
vanishing charge density was obtained in the geodesic approximation in
\cite{Keranen:2011xs}.} We then consider general scaling dimensions and solve
the scalar wave equation in the background spacetime of a charged Lifshitz black brane.

\subsection{Geodesic approximation}

First we consider operators with large scaling dimensions $\Delta$. On the gravitational
side this maps to large particle mass, for which we can approximate $m\approx \Delta$.
We can use the geodesic approximation to calculate the two point
function \cite{Balasubramanian:1999zv},
\beq
\langle\mathcal{O}(x)\mathcal{O}(x')\rangle \approx \epsilon^{-2\Delta}
e^{-\Delta\int d\tau\sqrt{g_{\mu\nu}\frac{dx^{\mu}}{d\tau}\frac{dx^{\nu}}{d\tau}}},
\label{eq:geodcor}
\eeq
where $x^{\mu}(\tau)$ is the geodesic of minimal length and $\epsilon$ is an infrared
cutoff in the bulk spacetime $r<1/\epsilon$.

The metrics of interest have the form (\ref{eq:metricansatz}) and we can parametrize theÊ
geodesics as $r=r(x)$,~$t=t(x)$ and $y=y(x)$, where $x$ and $y$ are transverse
coordinates. ÊBecause of translational symmetry in the $t$ and $y$ directions the
momenta  $p_t\propto t'(x)$ and $p_y\propto y'(x)$ are conserved.
By a combination of translation and rotation in the x-y plane we canÊalways
arrange both endpoints of the geodesic to be at $y=0$. It then follows that $y'(x)=0$
identically, since, if the derivative did not vanish, the geodesic could not start and end
at the same value of $y$. Similarly, for an equal time commutator, the constant value of
$t'(x)$ must in fact be zero for otherwise the two endpoints would be at different
values of $t$.

\begin{figure}[h]
\begin{center}
\includegraphics[scale=1.15]{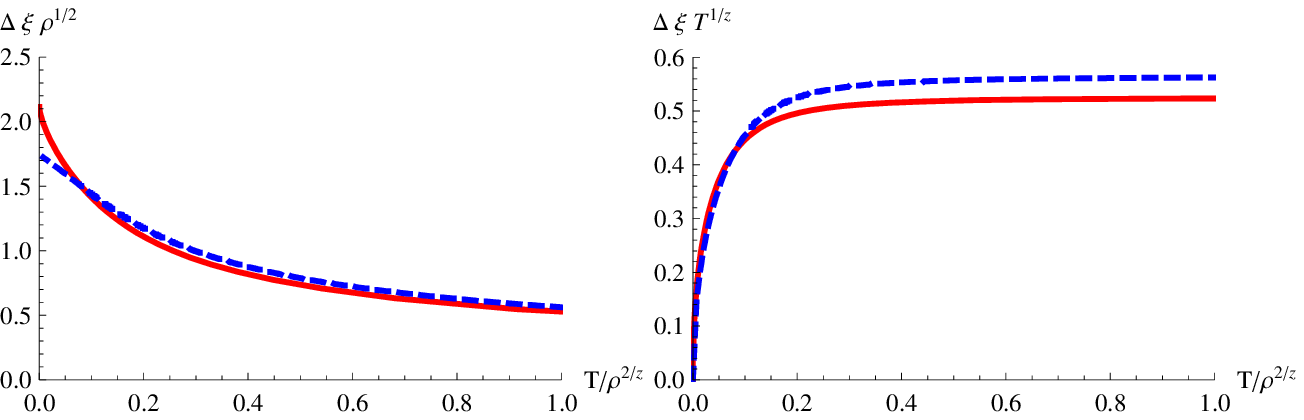}
\\
\includegraphics[scale=1.15]{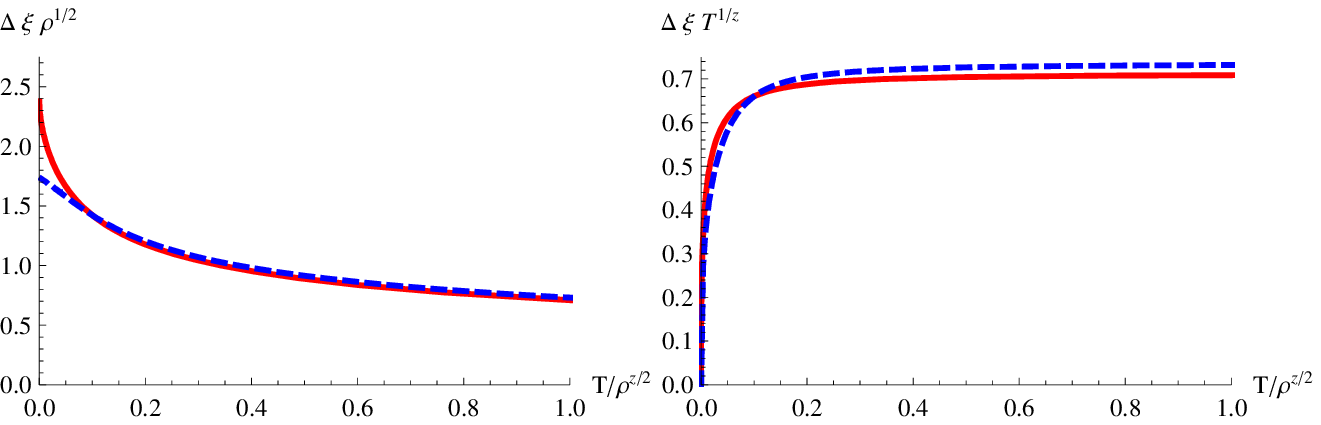}
\caption{\label{fig:geodcor} Correlation lengths in the two models as a function
of $T/\rho^{2/z}$ for the dynamical critical exponent $z=2$ (top) and $z=3$ (bottom).
The solid red curves correspond to the EMP model and the dashed blue curve
corresponds to the EDM model.}
\end{center}
\end{figure}

\begin{figure}[h]
\begin{center}
\includegraphics[scale=1]{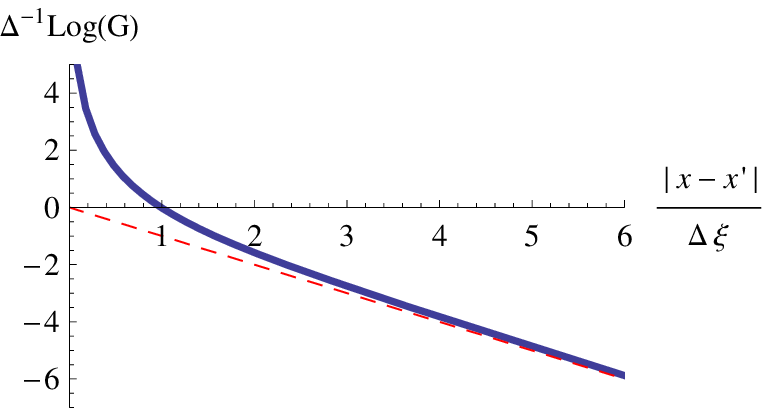}
\includegraphics[scale=1]{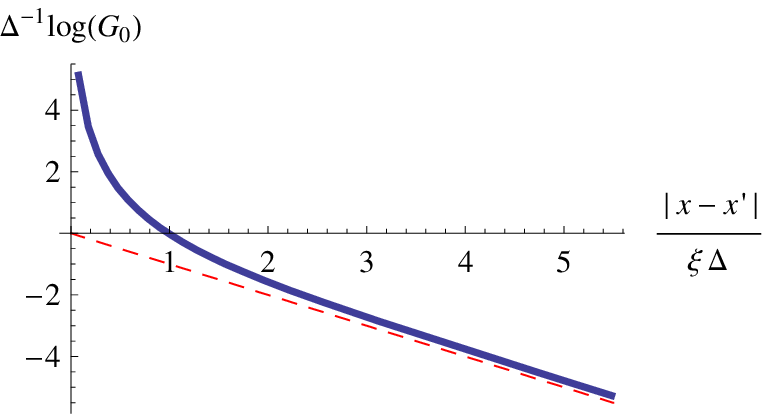}
\caption{\label{fig:logg} Logarithm of the two point equal time correlation function as a
function of distance for $z=2$. The dashed line corresponds to a reference line with
slope -1. The left figure is in the EDM model and $T/\rho^{z/2}\approx 0.0229$ while the
right figure is in the EMP model with $T/\rho^{z/2}\approx 0.109$. Correlators for
other temperature values are almost indistinguishable from the ones shown
in the figure. Most of the dependence on $z$ and $T$ comes through $\xi$.}
\end{center}
\end{figure}

In this case, the geodesic length functional simplifies,
\beq
S=\Delta\int dx\sqrt{r(x)^2+\frac{(r'(x))^2}{r(x)^2g(r)}}\equiv \Delta\int dxL.
\eeq
Since the ``Lagrangian" $L$ has no explicit $x$ dependence, there is a conserved
Hamiltonian,
\beq
H=\frac{\partial L}{\partial r'}r'-L=-\frac{r^2}{L}.
\eeq
The value of the Hamiltonian can be calculated at the turning point of the geodesic
at $r=r_*$. By translation symmetry the turning point can be taken to be at $x=0$ and then
we can use $r'(0)=0$ to give $H=-r_*$. This way we are lead to a first order differential
equation for the geodesic,
\beq
(r')^2=gr^4\left(\frac{r^2}{r^2_*}-1\right).\label{eq:geodeq}
\eeq
The on shell particle action is now given by
\beq
S=\Delta\int dr\frac{L}{r'}=\frac{2\Delta}{r_*}\int_{r_*}^{\epsilon^{-1}}dr
\frac{1}{\sqrt{g\Big(\frac{r^2}{r^2_*}-1\Big)}},
\label{eq:length}
\eeq
where we have used the chain rule and the equation of motion (\ref{eq:geodeq}).
It is straightforward to solve (\ref{eq:geodeq}) numerically and evaluate the integral
(\ref{eq:length}). Before doing this, we note that the two point function can be computed
analytically for large distances. First we write (\ref{eq:geodeq}) in integral form
\beq
\frac{1}{2}|\textbf{x}-\textbf{x}'|=\int_{r_*}^{\epsilon^{-1}}
dr\frac{1}{\sqrt{gr^4\Big(\frac{r^2}{r^2_*}-1\Big)}}.
\label{eq:length2}
\eeq
The key is to note that large distance $|\textbf{x}-\textbf{x}'|$ corresponds to a geodesic with
a turning point approaching the horizon $r=r_0$. In the limit $r_*\rightarrow r_0$ the
integrals (\ref{eq:length}) and (\ref{eq:length2}) both diverge logarithmically at the lower
end.\footnote{The integral in (\ref{eq:length}) is also divergent at the upper limit when the
cutoff is sent to zero. This divergence is independent of $r_*$ and has the same value in
the vacuum. We can ignore it as it cancels with the explicit power of $\epsilon$ in
(\ref{eq:geodcor}).}
So in this limit we can approximate
\beq
\frac{1}{2}|\textbf{x}-\textbf{x}'|=
\int_{r_*}^{\epsilon^{-1}}dr\frac{1}{\sqrt{gr^4\Big(\frac{r^2}{r^2_*}-1\Big)}}
\approx \frac{1}{r_*^2}\int_{r_*}dr\frac{1}{\sqrt{g\Big(\frac{r^2}{r^2_*}-1\Big)}}+...,
\eeq
where the dots denote terms that are subleading as $r_*\rightarrow r_0$. This way we get
\beq
S=\Delta r_0|\textbf{x}-\textbf{x}'|+...
\eeq
where the dots denote terms that increase slower than linearly in the limit of large
$|\textbf{x}-\textbf{x}'|$. Using (\ref{eq:geodcor}) we obtain
\beq
G(\textbf{x},\textbf{x}')=\langle\mathcal{O}(\textbf{x},t)\mathcal{O}(\textbf{x}',t)\rangle
\propto e^{-|\textbf{x}-\textbf{x}'|/\xi},
\eeq
where the correlation length is given by
\beq
\xi=\frac{1}{\Delta r_0}.
\label{eq:geodscreening}
\eeq
In order to obtain the temperature dependence of the correlation length we need
to know the relation between the temperature and the position of the horizon.
This has to be obtained mostly numerically. The temperature is related to the position
of the horizon through (\ref{eq:temperature}), keeping
$\rho$ in (\ref{eq:rho}) fixed.
Figure \ref{fig:geodcor} shows the results of a numerical evaluation of
(\ref{eq:geodscreening}) for both
holographic models as a function of $T/\rho^{z/2}$.
A numerical evaluation of the full correlation function in the geodesic approximation
for $z=2$ is shown in Figure \ref{fig:logg}. Correlation functions for other values
of $z$ and different values of $T/\rho^{z/2}$ are qualitatively very similar to the ones
shown in the figure. As can be seen from Figure \ref{fig:logg}, the numerical solution
indeed agrees with our (approximate) analytic result (\ref{eq:geodscreening}) at large
distances.

The short distance behavior of the correlation function has the form
\beq
G(\textbf{x},\textbf{x}')\approx\frac{1}{|\textbf{x}-\textbf{x}'|^{2\Delta}},
\eeq
independently of the temperature.

As can be seen from Figure \ref{fig:geodcor}, the correlation length approaches a finite
non-vanishing constant value as $T\rightarrow 0$ while keeping the charge density fixed.
For a vanishing chemical potential the correlation length diverges as $\xi\propto T^{-1/z}$
as Figure \ref{fig:geodcor} indicates.

\subsection{General scaling dimensions}

When the scaling dimensions of the operators in the two point function are small, the
geodesic approximation no longer applies and one has to solve the full wave-equation.
The Euclidean action for the boson is taken to have the standard form\footnote{We note that
the counter term in (\ref{eq:scalaraction}) is not the most general one, but it applies to
the values of the scaling dimension $\Delta<2+z/2$, to which we specialize in what follows. More generally one has terms
including derivatives with respect to the transverse dimensions \cite{Taylor:2008tg}.}
\beq
S_E=\frac{1}{2}\int d^4x\sqrt{-g}\Big((\partial\phi)^2+m^2\phi^2\Big)
+\frac{1}{2}\Delta_-\int d^3x\sqrt{-\gamma}\phi^2|_{r=\epsilon^{-1}},\label{eq:scalaraction}
\eeq
where $\gamma$ is the determinant of the induced metric at $r=\epsilon^{-1}$ and $\Delta_{-}$
is given below equation (\ref{eq:asymptotics}).
The two point function is obtained by solving the wave equation,
\beq
\frac{1}{\sqrt{-g}}\partial_{\mu}(\sqrt{-g}g^{\mu\nu}\partial_{\nu}\phi)-m^2\phi=0,
\eeq
in the bulk spacetime. We are interested in correlation functions at thermal equilibrium,
and thus we work in Euclidean signature. This means in particular that the
field $\phi$ should be Fourier transformed as
\beq
\phi(\textbf{x},r,\tau)=\sum_n\int\frac{d^2k}{(2\pi)^2}
e^{-i\omega_n\tau+i\textbf{k}\cdot\textbf{x}}\phi_n(r,k),
\eeq
where $\omega_n=2\pi T n$ are the Matsubara mode frequencies. To obtain the bulk
to boundary propagator we require regularity at the horizon. For the numerics we find it
convenient to use the coordinate $u=1/r$ in terms of which the metric reads
\beq
ds^2=f(u)\frac{d\tau^2}{u^{2z}}+\frac{1}{u^2}
\Big(\frac{du^2}{g(u)}+d\textbf{x}^2\Big).
\eeq

\begin{figure}[h]
\begin{center}
\includegraphics[scale=1]{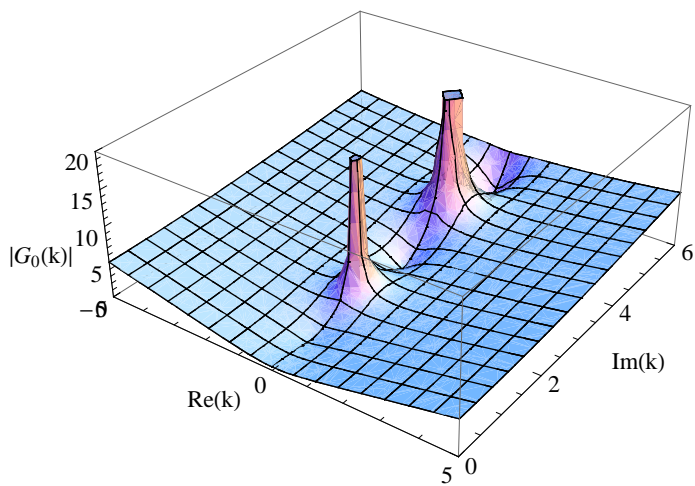}
\caption{\label{fig:matsubara} The momentum space correlator for $z=2$ and
$\omega_n=0$ in the EMD model at finite temperature and zero charge density.
The scaling dimension of the scalar operator is $\Delta=2+\sqrt{3}/2$.}
\end{center}
\end{figure}

We are lead to solve the wave equation
\beq
u^{3+z}\sqrt{\frac{g}{f}}\partial_u(u^{-1-z}\sqrt{fg}\partial_u\phi_n)-\Big(\frac{u^{2z}}{f}\omega_n^2+u^2k^2+m^2\Big)\phi_n=0,
\label{eq:scalarwave}
\eeq
together with the regularity condition, that, close to the horizon $u=u_0\equiv1/r_0$, the field
behaves as
\beq
\phi_n(u,k)\approx \phi_n^{(0)}
\exp{\left[-u_0^{z-1}\omega_n\int^{u}\frac{du'}{\sqrt{f(u')g(u')}}\right]},
\eeq
where $\phi_n^{(0)}$ has a regular Taylor expansion in $u-u_0$.
The modes $\phi_n(u,k)$ have the standard asymptotic behavior as $u\rightarrow 0$
\beq
\phi_n(u,k)=\phi_n^{(-)}u^{\Delta_-}+\phi_n^{(+)}u^{\Delta_+}+...,\label{eq:asymptotics}
\eeq
where $\Delta_{\pm}=(2+z)/2\pm \nu$ and $\nu=\sqrt{(2+z)^2/4+m^2}$.
The holographic dictionary with our Fourier transform conventions reads
\beq
e^{-S_E}=\left\langle \exp{\left[\beta\sum_n\int\frac{d^2k}{(2\pi)^2}\mathcal{O}_n(k)
\phi_{-n}^{(-)}(-k)\right]}\right\rangle,
\eeq
which leads to the two point function
\beqa
\langle\mathcal{O}_n(k)\mathcal{O}_{n'}(k')\rangle&=&
-\frac{(2\pi)^4}{\beta^2}\frac{\delta^2 S_E}{\delta\phi_{-n}^{(-)}(-k)\delta\phi_{-n'}^{(-)}(-k')} \\
&=&T\delta_{n+n',0}\delta^2(k+k')(2\pi)^2G_n(k),\nonumber
\eeqa
where
\beq
G_n(k)=2\nu\frac{\phi^{(+)}_n}{\phi^{(-)}_n}.\label{eq:wavecorr}
\eeq
For the alternative quantization \cite{Balasubramanian:1998sn,Klebanov:1999tb} (which corresponds to treating the subleading mode $\phi^{(+)}$ as the source) we have
\beq
G_n(k)=-(2\nu)^{-1}\frac{\phi^{(-)}_n}{\phi^{(+)}_n}.\label{eq:alternative}
\eeq
The momentum space correlator has poles on the imaginary $k$ axis, which means
that the real space two point function decays exponentially,
$G(\textbf{x},\textbf{x}')\propto e^{-k_*|\textbf{x}-\textbf{x}'| }$, where $k_*$ is the pole
closest to the real $k$ axis. It is sufficient to take into account the lowest Matsubara mode
as the higher ones have a faster fall of in position space\footnote{The same procedure
has been used earlier in calculating the Debye screening length using holography
in $\mathcal{N}=2^*$ supersymmetric Yang-Mills theory \cite{Hoyos:2011uh}.}.
We solve (\ref{eq:scalarwave}) numerically and extract the momentum space two
point correlator using (\ref{eq:wavecorr}), with $n=0$. Results from a numerical
evaluation of the finite temperature two point correlator is shown in
Figure \ref{fig:matsubara}. This indeed shows poles on the imaginary $k$ axis.

\begin{figure}[h]
\begin{center}
\includegraphics[scale=1.2]{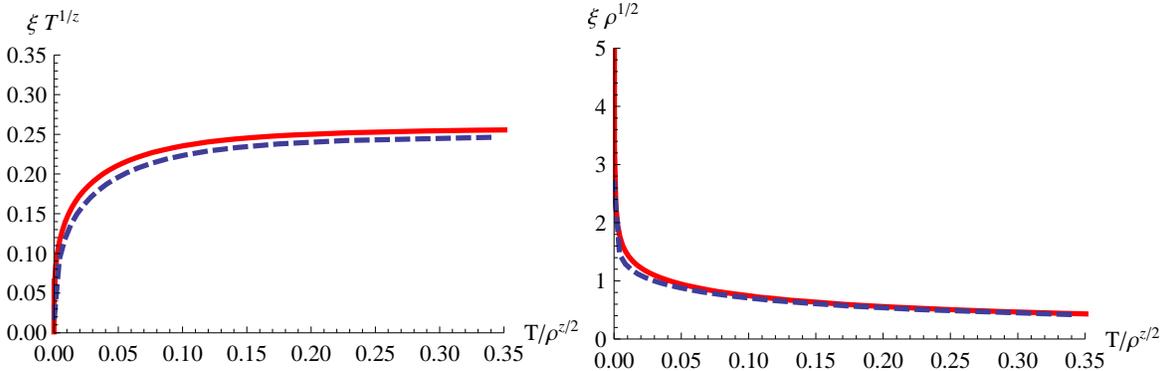}
\caption{\label{fig:matsubara2} The correlation length, as extracted from the position
of the lowest pole of the momentum space correlator, as a function of $T/\rho^{z/2}$.
The dashed (blue) curves are in the EMD model while the solid (red) curves are in the
EMP model. All the plots are for $z=2$ and $\Delta=2+\sqrt{3}/2$.}
\end{center}
\end{figure}

By tracking the position of the lowest pole in the momentum space propagator, we can
obtain the correlation length as a function of $T/\rho^{z/2}$ as the temperature is varied.
The correlation length for $z=2$ obtained this way is shown in Figure \ref{fig:matsubara2}
for both holographic models.

\section{Double trace deformations}\label{sec:doubletracecorr}
In the previous section we considered the finite temperature dependence of
scalar correlation functions at finite charge density. While working at finite charge
density gives rise to non-trivial dependence on the dimensionless ratio of the
length scales associated to $\rho$ and $T$, it has the drawback that at zero
temperature there is still a finite length scale in the problem, associated with the
charge density.

Another way to study the temperature dependence of the correlation length
is to work at zero charge density and add a double trace deformation
for the scalar operator \cite{Berkooz:2002ug,Witten:2001ua},
\beq
\delta S_{E}=\frac{1}{2}\int d^3x \lambda\mathcal{O}(x)^2.
\eeq
The coupling constant $\lambda$ introduces a new reference length scale into the
problem, which, when combined with $T$, gives rise to a dimensionless ratio
that enters into thermal correlators. For a concise review
of double trace deformations see the Appendix of \cite{Faulkner:2010jy}.

A prescription for calculating two point correlation functions in AdS/CFT in the presence of a
double trace deformation was given in \cite{Mueck:2002gm}. The same argument goes through
in asymptotically Lifshitz spacetime and we do not repeat it here.
Denoting the original momentum space two point function as $G_n(k)$, the two
point function in the deformed theory is found to be
\beq
G_n^{(\lambda)}(k)=\frac{G_n(k)}{1+\lambda G_n(k)}.
\label{eq:dt}
\eeq
The form of the two point function (\ref{eq:dt}) can also be understood in a simple way from
large-$N$ factorization of correlation functions. The deformed two point function is
\beq
\left\langle \mathcal{O}_n(k)\mathcal{O}_{n'}(k')
e^{-\frac{1}{2}\beta\sum_n\int \frac{d^2q}{(2\pi)^2}
\lambda\mathcal{O}_n(q)\mathcal{O}_{-n}(-q)}\right\rangle
=T\delta_{n+n',0}\delta^2(k+k')(2\pi)^2G_n^{(\lambda)}(k),
\eeq
which after expanding the exponential in a power series, and using large-N factorization,
becomes
\beq
G_n^{(\lambda)}(k)=\sum_{m=1}^{\infty}G_n(k)^m\lambda^{m-1}
=\frac{G_n(k)}{1+\lambda G_n(k)},
\eeq
recovering (\ref{eq:dt}).

\begin{figure}[h]
\begin{center}
\includegraphics[scale=1.2]{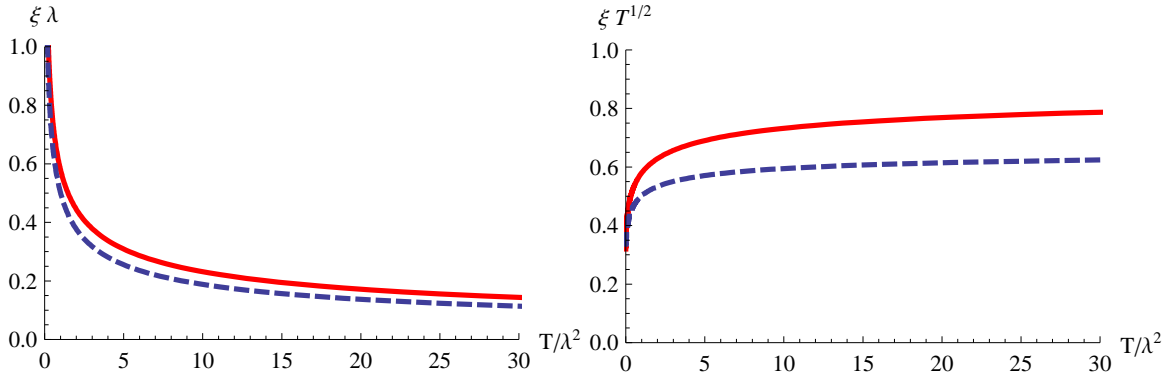}
\caption{\label{fig:dt} The correlation length as extracted from the position of the lowest
pole of the momentum space correlator in the double trace deformed theory. The solid
red curves are in the EMP theory while the dashed blue curves are in the EDM theory.
Both plots are for $z=2$ and $\Delta=3/2$.}
\end{center}
\end{figure}

The method that was described in the previous section for solving for the two point
function $G(k)$ applies here as well. Since we are interested in relevant deformations
of the theory, we consider correlation functions in the alternative
quantization \cite{Balasubramanian:1998sn,Klebanov:1999tb}, for which the
undeformed correlator is given by (\ref{eq:alternative}). In this case the deformed
correlator becomes
\beq
G_n^{(\lambda)}(k)=
\Big(\lambda-2\nu \frac{\phi^{(+)}_n(k)}{\phi^{(-)}_n(k)}\Big)^{-1}.
\label{eq:dt2}
\eeq
By tracking the lowest pole of the deformed two point function (\ref{eq:dt2}) we obtain
the correlation length shown in Figure \ref{fig:dt}.
At high temperature the dimensionless correlation length $\xi T^{1/2}$ is simply that of
the alternative quantization without a double trace deformation. As the temperature is
lowered, the dimensionless combination $\xi T^{1/2}$ flows from its value in the alternative quantization, to its value in the standard quantization. This reflects the renormalization
group flow from the alternative quantization to the standard
one \cite{Klebanov:1999tb,Faulkner:2010jy}. In the vacuum the renormalization group
flow is seen when considering the behavior of the correlation function as a function of
distance, short distances corresponding to the alternative quantization and large
distances to the standard quantization. In our case the renormalization group flow is
instead revealed as a function of the temperature.

\section{Discussion}\label{sec:conclusions}

In this paper we have considered two bottom up holographic models both of which admit
a metric with a Lifshitz scaling symmetry, with arbitrary dynamical critical exponent, and
include a $U(1)$ gauge field. Both models have charged black brane solutions that
give access to finite temperature physics in the corresponding dual field theories. At high
temperature (compared to the scale set the $U(1)$ charge density) the behavior of
thermodynamic quantities is governed by the underlying Lifshitz scaling symmetry.
As the temperature is lowered, keeping the charge density fixed, various thermodynamic
quantities in the two models start to differ from each other. In particular, a sharp distinction
between the two models appears in the specific heat, which for the EMD model behaves
at low temperatures as
\beq
\frac {C}{T}\rightarrow\textrm{const.} \quad \textrm{as} \quad T\rightarrow 0,
\eeq
while numerical results suggest that $C/T$ grows without bound in the limit of
zero temperature in the EMP model \cite{Brynjolfsson:2010rx}.
It thus appears that the same relevant deformation, by a chemical potential, takes us to
a different kind of a fixed point in the zero temperature limit in the two models, at least
as far as thermodynamics is concerned.
The infrared (near horizon) geometry of extremal black branes has an
$\textrm{AdS}_2\times\textrm{R}^2$ region in both of the models, as discussed in
section~\ref{sec:hololifs}, but in the EMP model only after a singular coordinate
transformation.

For the finite temperature equal time two point function, at finite charge density,
we find the standard power law behavior at short distances, while for large
distances it decays exponentially due to the presence of a horizon in the bulk.
The correlation length extracted from the exponential decay is shown in
Figure~\ref{fig:geodcor} for large scaling dimension operators and in
Figure~\ref{fig:matsubara2} for generic scaling dimensions. For high
temperatures $T/\rho^{z/2}> 0.2$, the temperature dependence of the correlation
length is consistent with scale invariance, implying a large temperature range
where the critical point dominates the physics.
As expected, the two models agree quite well in this range. On the other hand,
the correlation length exhibits distinctly different behavior in the two models
at low temperature and fixed charge density. In the EMD model we obtain the
following behavior in the geodesic approximation,
\beq
\Big(\frac{\partial \xi}{\partial T}\Big)_{\rho}\rightarrow\textrm{const.},
\eeq
while the same quantity diverges in the zero temperature limit in the EMP model,
\beq
\Big(\frac{\partial \xi}{\partial T}\Big)_{\rho}\rightarrow -\infty.
\eeq
In both cases $\xi$ tends to a finite value as $T\rightarrow 0$ when the charge density
is kept fixed at a non-zero value.

The correlation length in the presence of a double trace deformation is shown in
Figure~\ref{fig:dt}. In this case the correlation length diverges in both models in the
zero temperature limit, while keeping the double trace coupling $\lambda$ fixed.
The temperature dependence of the correlation length has a scale invariant form
in two regions, at high temperatures and at low temperatures. The combination
$\xi T^{1/z}$ runs between its fixed point values, from the alternative quantization
at high temperatures to the standard quantization at low temperatures. Overall,
the double trace deformation leads to a qualitatively very similar behavior in both
holographic models.

The double trace deformation leads to a Lifshitz symmetric result in the zero
temperature limit while at fixed charge density there remains a finite length
scale in the problem even at zero temperature. We can therefore more meaningfully
compare the double trace deformation case to the quantum Lifshitz model, which, as
was reviewed in section~\ref{sec:quantumlifs}, has several temperature dependent
length scales. The scale associated with a vortex plasma in the quantum Lifshitz
model behaves as $\xi_{vortex}\propto e^{-E_{c}/2T}$, which changes more rapidly
with $T$ than we find in the holographic correlation functions. The comparison to the
``intermediate" length scale $\xi_T$ in (\ref{eq:QLscale}), given by
\beq
\xi_T T^{1/2}\propto\sqrt{\frac{\log(-\log T)}{-\log T}},
\eeq
works somewhat better. The dimensionless combination $\xi_T T^{1/2}$ is seen to
be a fairly mild function of temperature, even if it diverges in the zero temperature limit.
At a qualitative level, the double trace deformation gives rise to similar behavior as it
induces a mild temperature dependence on $\xi T^{1/z}$.

Both types of deformations in both holographic models show similar behavior for the
combination $\xi T^{1/z}$. It is seen to be an increasing function of temperature in
all of the cases we have studied here. It would be interesting to understand this from
the perspective of the finite temperature renormalization group.

\section{Acknowledgements}

This work was supported in part by the Icelandic Research Fund and
by the University of Iceland Research Fund. We would like to thank Tobias Zingg and
Sean Nowling for discussions.


\begin{thebibliography}{99}

\bibitem{Hartnoll:2009sz} S.~A.~Hartnoll, ``Lectures on holographic methods
for condensed matter physics,'' Class.\ Quant.\ Grav.\  {\bf 26}, 224002 (2009)
[arXiv:0903.3246 [hep-th]].

\bibitem{Herzog:2009xv}C.~P.~Herzog, ``Lectures on Holographic Superfluidity
and Superconductivity,'' J.\ Phys.\ A  {\bf 42}, 343001 (2009) [arXiv:0904.1975 [hep-th]].

\bibitem{Horowitz:2010gk}G.~T.~Horowitz, ``Introduction to Holographic
Superconductors,'' arXiv:1002.1722 [hep-th].

\bibitem{McGreevy:2009xe}
  J.~McGreevy,
  ``Holographic duality with a view toward many-body physics,''
  Adv.\ High Energy Phys.\  {\bf 2010}, 723105 (2010)
  [arXiv:0909.0518 [hep-th]].

\bibitem{Sachdev:2010ch}S.~Sachdev, ``Condensed matter and AdS/CFT,''
arXiv:1002.2947 [hep-th].

\bibitem{Sachdev}
  S.~Sachdev,
  ``Quantum phase transitions,"
  Cambridge U. Press (2000).

\bibitem{Sondhi}
  S.~L.~Sondhi, S.~M.~Girvin, J.~P.~Carini, D.~Shahar
  ``Continuous quantum phase transitions,"
  Rev.\ Mod.\ Phys.\ {\bf 69}, (1997) 315–333.
  [cond-mat/9609279].

\bibitem {Kachru:2008yh}S.~Kachru, X.~Liu, and M.~Mulligan, ``Gravity Duals of
Lifshitz-like Fixed Points," Phys.\ Rev.\ D \textbf{78} (2008) 106005
[arXiv:0808.1725 [hep-th]].

\bibitem{Taylor:2008tg}M.~Taylor,
``Non-relativistic holography,''
arXiv:0812.0530 [hep-th].

\bibitem{Brynjolfsson:2009ct}
  E.~J.~Brynjolfsson, U.~H.~Danielsson, L.~Thorlacius and T.~Zingg,
  ``Holographic Superconductors with Lifshitz Scaling,''
  J.\ Phys.\ A A {\bf 43}, 065401 (2010)
  [arXiv:0908.2611 [hep-th]].

\bibitem{Tarrio:2011de}
  J.~Tarrio and S.~Vandoren,
  ``Black holes and black branes in Lifshitz spacetimes,''
  JHEP {\bf 1109}, 017 (2011)
  [arXiv:1105.6335 [hep-th]].

\bibitem{Viswanath}
  A.~Viswanath, L.~Balents, T.~Senthil,
  ``Quantum criticality and deconfinement in phase transitions between valence bond solids,''
  Phys.\ Rev.\  {\bf B69 } (2004)  224416.
  [cond-mat/0311085].

\bibitem{Fradkin}
  E.~Fradkin, D.~A.~Huse, R.~Moessner, V.~Oganesyan, S.~L.~Sondhi,
  ``Bipartite Rokhsar–Kivelson points and Cantor deconfinement,''
  Phys.\ Rev.\ {\bf B69} (2004) 224415.
  [cond-mat/0311353].

\bibitem{Ardonne}
  E.~Ardonne, P.~Fendley, E.~Fradkin,
  ``Topological Order and Conformal Quantum Critical Points,''
  Ann.\ Phys.\ (N.Y.) {\bf 310} (2004) 493.
  [cond-mat/0311466].

\bibitem{Ghaemi}
 P.~Ghaemi, A.~Vishwanath, T.~Senthil,
 ``Finite-temperature properties of quantum Lifshitz transitions between valence-bond solid phases: An example of local quantum criticality,''
  Phys.\ Rev.\  {\bf B72 } (2005)  024420.
  [cond-mat/0412409].

\bibitem{Koroteev:2007yp}
 P.~Koroteev, M.~Libanov,
 ``On Existence of Self-Tuning Solutions in Static Braneworlds without Singularities,''
  JHEP {\bf 0802}, 104 (2008).  [arXiv:0712.1136 [hep-th]].

\bibitem{Brynjolfsson:2010mk}
  E.~J.~Brynjolfsson, U.~H.~Danielsson, L.~Thorlacius and T.~Zingg,
  ``Holographic Models with Anisotropic Scaling,''
  arXiv:1004.5566 [hep-th].

\bibitem{Pang:2009pd}
  D.~-W.~Pang,
  ``On Charged Lifshitz Black Holes,''
  JHEP {\bf 1001}, 116 (2010)
  [arXiv:0911.2777 [hep-th]].

\bibitem{Danielsson:2009gi}
  U.~H.~Danielsson and L.~Thorlacius,
  ``Black holes in asymptotically Lifshitz spacetime,''
  JHEP {\bf 0903}, 070 (2009)
  [arXiv:0812.5088 [hep-th]].

\bibitem{Keranen:2011xs}
  V.~Keranen, E.~Keski-Vakkuri and L.~Thorlacius,
  ``Thermalization and entanglement following a non-relativistic holographic quench,''
  Phys.\ Rev.\ D {\bf 85}, 026005 (2012)
  [arXiv:1110.5035 [hep-th]].

\bibitem{Bertoldi:2009vn}
  G.~Bertoldi, B.~A.~Burrington and A.~Peet,
  ``Black Holes in asymptotically Lifshitz spacetimes with arbitrary critical exponent,''
  Phys.\ Rev.\ D {\bf 80}, 126003 (2009)
  [arXiv:0905.3183 [hep-th]].

\bibitem{Ross:2009ar}
  S.~F.~Ross and O.~Saremi,
  ``Holographic stress tensor for non-relativistic theories,''
  JHEP {\bf 0909}, 009 (2009)
  [arXiv:0907.1846 [hep-th]].

\bibitem{Zingg:2011cw}
  T.~Zingg,
  ``Thermodynamics of Dyonic Lifshitz Black Holes,''
  JHEP {\bf 1109}, 067 (2011)
  [arXiv:1107.3117 [hep-th]].

\bibitem{Ross:2011gu}
  S.~F.~Ross,
  ``Holography for asymptotically locally Lifshitz spacetimes,''
  Class.\ Quant.\ Grav.\  {\bf 28}, 215019 (2011)
  [arXiv:1107.4451 [hep-th]].

\bibitem{Baggio:2011cp}
  M.~Baggio, J.~de Boer and K.~Holsheimer,
  ``Hamilton-Jacobi Renormalization for Lifshitz Spacetime,''
  JHEP {\bf 1201} (2012) 058
  [arXiv:1107.5562 [hep-th]].

\bibitem{Mann:2011hg}
  R.~B.~Mann and R.~McNees,
  ``Holographic Renormalization for Asymptotically Lifshitz Spacetimes,''
  JHEP {\bf 1110}, 129 (2011)
  [arXiv:1107.5792 [hep-th]].

\bibitem{Tarrio:2012xx}
  J.~Tarrio,
  ``Asymptotically Lifshitz Black Holes in Einstein-Maxwell-Dilaton Theories,''
  arXiv:1201.5480 [hep-th].

\bibitem{Cubrovic:2009ye}
  M.~Cubrovic, J.~Zaanen and K.~Schalm,
  ``String Theory, Quantum Phase Transitions and the Emergent Fermi-Liquid,''
  Science {\bf 325} (2009) 439
  [arXiv:0904.1993 [hep-th]].

\bibitem{Liu:2009dm}
  H.~Liu, J.~McGreevy and D.~Vegh,
  ``Non-Fermi liquids from holography,''
  Phys.\ Rev.\ D {\bf 83} (2011) 065029
  [arXiv:0903.2477 [hep-th]].

\bibitem{Faulkner:2009wj}
  T.~Faulkner, H.~Liu, J.~McGreevy and D.~Vegh,
  ``Emergent quantum criticality, Fermi surfaces, and AdS(2),''
  Phys.\ Rev.\ D {\bf 83} (2011) 125002
  [arXiv:0907.2694 [hep-th]].

\bibitem{Gursoy:2011gz}
  U.~Gursoy, E.~Plauschinn, H.~Stoof and S.~Vandoren,
  ``Holography and ARPES Sum-Rules,''
  arXiv:1112.5074 [hep-th].

\bibitem{Fang:2012pw}
  L.~Q.~Fang, X.~-H.~Ge and X.~-M.~Kuang,
  ``Holographic fermions in charged Lifshitz theory,''
  arXiv:1201.3832 [hep-th].

\bibitem{Alishahiha:2012nm}
  M.~Alishahiha, M.~R.~Mohammadi Mozaffar and A.~Mollabashi,
  ``Fermions on Lifshitz Background,''
  arXiv:1201.1764 [hep-th].

\bibitem{Brynjolfsson:2010rx}
  E.~J.~Brynjolfsson, U.~H.~Danielsson, L.~Thorlacius and T.~Zingg,
  ``Black Hole Thermodynamics and Heavy Fermion Metals,''
  JHEP {\bf 1008} (2010) 027
  [arXiv:1003.5361 [hep-th]].

\bibitem{Bertoldi:2009dt}
  G.~Bertoldi, B.~A.~Burrington and A.~W.~Peet,
  ``Thermodynamics of black branes in asymptotically Lifshitz spacetimes,''
  Phys.\ Rev.\ D {\bf 80}, 126004 (2009)
  [arXiv:0907.4755 [hep-th]].

\bibitem{Balasubramanian:2009rx}
  K.~Balasubramanian and J.~McGreevy,
  ``An Analytic Lifshitz black hole,''
  Phys.\ Rev.\ D {\bf 80} (2009) 104039
  [arXiv:0909.0263 [hep-th]].

\bibitem{Giacomini:2012hg}
  A.~Giacomini, G.~Giribet, M.~Leston, J.~Oliva and S.~Ray,
  ``Scalar field perturbations in asymptotically Lifshitz black holes,''
  arXiv:1203.0582 [hep-th].

\bibitem{Balasubramanian:1999zv}
  V.~Balasubramanian, S.~F.~Ross,
  ``Holographic particle detection,''
  Phys.\ Rev.\  {\bf D61 } (2000)  044007.
  [hep-th/9906226].

\bibitem{Balasubramanian:1998sn}
  V.~Balasubramanian, P.~Kraus and A.~E.~Lawrence,
  ``Bulk versus boundary dynamics in anti-de Sitter space-time,''
  Phys.\ Rev.\ D {\bf 59} (1999) 046003
  [hep-th/9805171].

\bibitem{Klebanov:1999tb}
  I.~R.~Klebanov and E.~Witten,
  ``AdS / CFT correspondence and symmetry breaking,''
  Nucl.\ Phys.\ B {\bf 556} (1999) 89
  [hep-th/9905104].

\bibitem{Hoyos:2011uh}
  C.~Hoyos, S.~Paik and L.~G.~Yaffe,
  ``Screening in strongly coupled N=2* supersymmetric Yang-Mills plasma,''
  JHEP {\bf 1110} (2011) 062
  [arXiv:1108.2053 [hep-th]].

\bibitem{Berkooz:2002ug}
  M.~Berkooz, A.~Sever and A.~Shomer,
  ``'Double trace' deformations, boundary conditions and space-time singularities,''
  JHEP {\bf 0205} (2002) 034
  [hep-th/0112264].

\bibitem{Witten:2001ua}
  E.~Witten,
  ``Multitrace operators, boundary conditions, and AdS / CFT correspondence,''
  hep-th/0112258.

\bibitem{Faulkner:2010jy}
  T.~Faulkner, H.~Liu and M.~Rangamani,
  ``Integrating out geometry: Holographic Wilsonian RG and the membrane paradigm,''
  JHEP {\bf 1108} (2011) 051
  [arXiv:1010.4036 [hep-th]].

\bibitem{Mueck:2002gm}
  W.~Mueck,
  ``An Improved correspondence formula for AdS / CFT with multitrace operators,''
  Phys.\ Lett.\ B {\bf 531} (2002) 301
  [hep-th/0201100].

\end{thebibliography}
\end{document}